\documentclass{ws-procs9x6}

\usepackage[british]{babel}
\usepackage[T1]{fontenc}

\usepackage{citesort}                     
\usepackage[dvips]{color}                 
\usepackage{graphicx}              

\usepackage{multirow}                     
\usepackage{amssymb}
\usepackage{amsmath}
\usepackage{xspace}                       

\newcommand{\NPA}{\textnormal{Nucl.\ Phys.\ }\textnormal{A}}

\newcommand{\PLB}{\textnormal{Phys.\ Lett.\ }\textnormal{B}}
\newcommand{\PR}{\textnormal{Phys.\ Rev.\ }}

\newcommand{\PRC}{\PR\textnormal{C}~}

\newcommand{\MeV}{\ensuremath{\mathrm{MeV}}}
\newcommand{\fm}{\ensuremath{\mathrm{fm}}}
\newcommand{\ChiEFT}{$\chi$EFT\xspace}

 \newcommand{\calO}{\mathcal{O}}

\begin{document}

\title{Nucleon Polarisabilities from Deuteron Compton Scattering, and Its
  Lessons for Chiral Power Counting}

\author{Harald W.~Grie\3hammer$^*$\footnote{Preprint nucl-th/0611074.
    Contribution to the \textsc{5th International Workshop on Chiral Dynamics,
      Theory and Experiment}, Durham/Chapel Hill NC (USA), 18th -- 22nd
    September 2006. To be published in the proceedings.}}

\address{Center for Nuclear Studies, Department of Physics,\\
  The George Washington University, Washington DC, USA.  $^*$E-mail:
  hgrie@gwu.edu}



\bodymatter

Polarisabilities measure the global stiffness of the nucleon's constituents
against displacement in an external electro-magnetic field. We examined them
in elastic deuteron Compton scattering $\gamma d\to\gamma d$ for photon
energies between zero and $130\;\MeV$ in Chiral Effective Field Theory \ChiEFT
with explicit $\Delta(1232)$ degrees of freedom, see
Refs.~\cite{polasfromdeuteron2,HildebrandtPhD} for details and better
references.  An excellent tool to identify the active low-energy degrees of
freedom are the \emph{dynamical polarisabilities}, defined by a multipole
decomposition of the structure part of the Compton amplitude at fixed energy.
Unique signals allow one to study the temporal response of each constituent.

\begin{figure}
 \includegraphics*[width=0.463\linewidth]{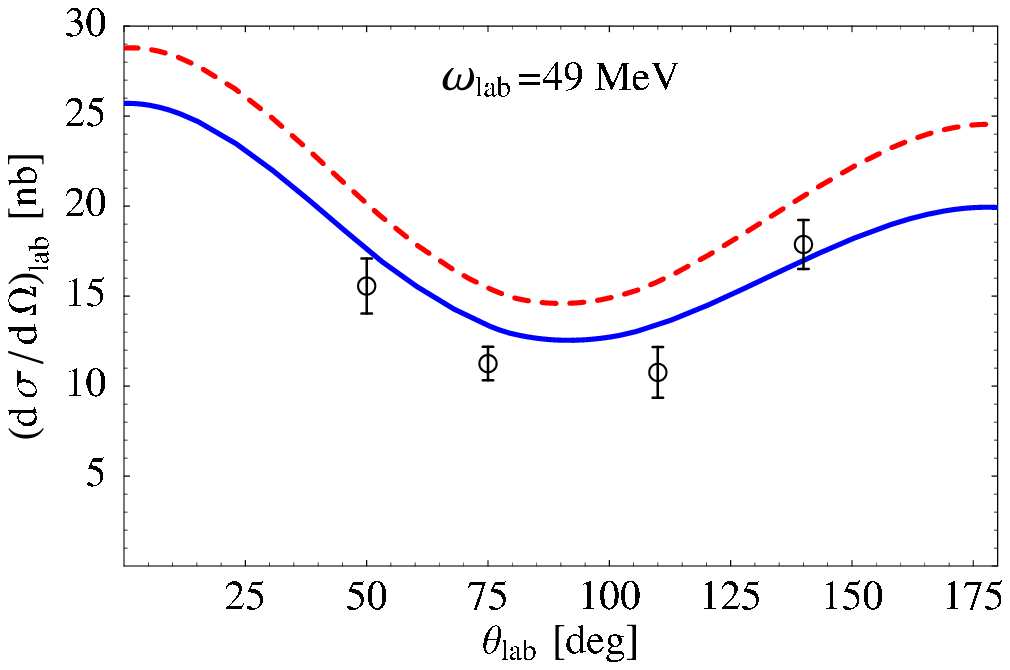}
 \hfill
 \includegraphics*[width=0.48\linewidth]{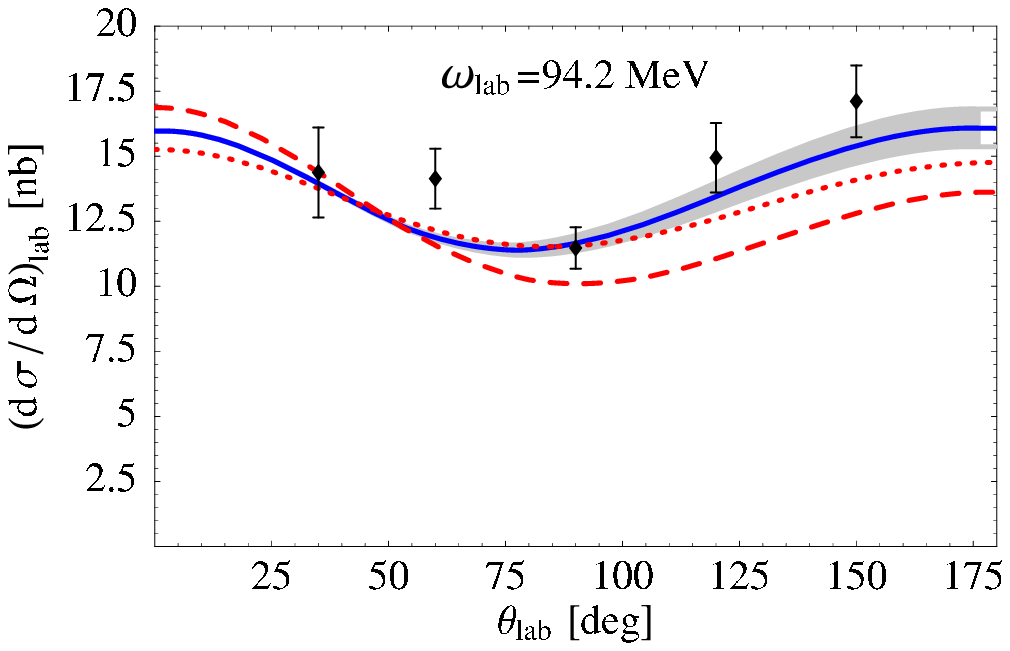}

 \caption{Left: Example of prediction using proton polarisabilities with
   (solid) and without (dashed) $NN$-rescattering in intermediate states.
   Right: Example of 1-parameter fit result using the Baldin sum rule for the
   deuteron (solid, with stat.~uncertainty), compared to \ChiEFT without
   explicit $\Delta(1232)$ ($\calO(p^3)$, dashed) and to a fit~\cite{judith}
   at $\calO(p^4)$ ($\alpha^s_{E1}=11.5,\;\beta^s_{M1}=0.3$, dotted).  From
   Ref.~\cite{polasfromdeuteron2}.}
\label{fig:1}
\end{figure}

For example, the strong, para-magnetic $N$-to-$\Delta(1232)$ transition
induces a strong energy-dependence which is pivotal to resolve the
``SAL-puzzle'', see Fig.~\ref{fig:1}: While all previous analyses of the
SAL-data at $95\;\MeV$ extracted vastly varying nucleon polarisabilities,
\ChiEFT with an explicit $\Delta(1232)$ captures correctly both normalisation
and angular dependence of the data without altering the static (namely
zero-energy) polarisabilities.

A consistent description must also give the correct Thomson limit, i.e.~the
exact low-energy theorem which is a consequence of gauge invariance. Its
verification is straight-forward in the 1-nucleon sector, where the amplitude
is perturbative. In contradistinction, all terms of the leading-order
Lippmann-Schwinger equation of $NN$-scattering, including the potential (and
hence one-pion exchange), must be of order $Q^{-1}$ when all nucleons are
close to their non-relativistic mass-shell, to accomodate the shallow
bound-state~\cite{hg}. In a consistent power-counting, all $NN$-rescattering
processes between photon-absorption and emission must thus be included.  Our
Green's function approach embeds these diagrams to guarantee the Thomson
limit, which is however statistically significant only below $70\;\MeV$, see
Fig.~\ref{fig:1}.  Up to next-to-leading order, the only unknowns are
contributions to the polarisabilities from short-distance Physics, leading to
two energy-independent parameters. The iso-scalar Baldin sum rule is in
excellent agreement with our 2-parameter fit, serving as input to
model-independently determine the iso-scalar, spin-independent dipole
polarisabilities of the nucleon at zero energy from all Compton data below
$100\;\MeV$:
\[
  \alpha_{E1}^s=11.3\pm0.7_\mathrm{stat}\pm0.6_\mathrm{Baldin}\pm1_\mathrm{th}\;,\; 
  \beta_{M1}^s =3.2\mp0.7_\mathrm{stat}\pm0.6_\mathrm{Baldin}\pm1_\mathrm{th}
\] 
(in $10^{-4}\;\fm^3$). We estimate the theoretical uncertainty to be $\pm1$
from typical higher-order contributions in the 1- and 2-nucleon sector.
Dependence on the $NN$-potential or deuteron wave-function used is virtually
eliminated with the correct Thomson limit.  Comparing this with our analysis
of all proton Compton data below $170\;\MeV$ by the same method,
\[
  \alpha_{E1}^p=11.0\pm1.4_\mathrm{stat}\pm0.4_\mathrm{Baldin}\pm1_\mathrm{th}\;,\;
  \beta_{M1}^p =2.8\mp1.4_\mathrm{stat}\pm0.4_\mathrm{Baldin}\pm1_\mathrm{th}\;,
\]
we conclude that the proton and neutron polarisabilities are to this leading
order identical within (predominantly statistical) errors, as predicted by
\ChiEFT. More and better data from MAXlab (Lund) will lead to a more precise
extraction, allowing one to zoom in on the proton-neutron differences.

\vspace*{-1.1ex}


\end{document}